\documentstyle[12pt]{article}
\begin{document}
\begin{titlepage}

\begin{flushright}
INS-Rep.-1012\\
NDA-FP-13/93\ \\
OCHA-PP-37\ \ \ \\
October 1993
\end{flushright}

\vfill

\begin{center}

{\Large\bf Wormholes in Two Dimensional Dilaton Gravity}

\vfill

\bigskip{Shin'ichi {\sc Nojiri}\footnote{
NOJIRI@JPNKEKVX, KEKVAX::NOJIRI}
\\
{\it Department of Mathematics and Physics\\
National Defense Academy\\
Hashirimizu, Yokosuka 239, JAPAN}}

\vfill

\bigskip{Ichiro {\sc Oda}\\
{\it Institute for Nuclear Study, University of Tokyo\\
3-2-1 Midori-cho, Tanashi-shi, Tokyo 188, JAPAN}\\
and\\
{\it Faculty of Science, Department of Physics\\
Ochanomizu University\\
1-1, Otsuka 2, Bunkyo-ku, Tokyo 112, JAPAN}}

\end{center}

\vfill

\begin{abstract}
It is shown that the general solution of classical equations of motion
in two dimensional dilaton gravity proposed
by Callan, Giddings, Harvey and Strominger (CGHS)
includes a Lorentzian wormhole solution in addition to a black hole solution.
We also show that matter perturbation of the wormhole by a shock wave
leads to the formation of a black hole where the curvature
singularity is cloaked by the global event horizon.
It is also argued that the classical
wormhole would be stable against quantum corrections.
\end{abstract}

\end{titlepage}

Semi-classical calculations, where a black hole is regarded as
a fixed background, tells that the radiation from
the black hole is thermally distributed just as one might expect
from an ideal black body \cite{HawkingI}.
The semi-classical calculation would be justified
when the mass is much larger than the Planck mass.
If the above result is true even in the full quantum theory,
we cannot construct a quantum theory of gravity in the ordinary
framework of the quantum mechanics.

Let us suppose a situation where a black hole is formed by
the accumulation of matters.
The corresponding initial state can be regarded as a pure quantum state.
Just after a black hole is formed, there begins the thermal radiation,
which corresponds to a mixed quantum state.
If the scenario by Hawking is true, the black hole is
expected to evaporate completely without leaving any remnant behind.
The evolution of a pure quantum state into a mixed one, however,
certainly violates the fundamental principles of quantum mechanics
such as unitarity and CPT invariance \cite{Page}.

Since we have not found any renormalizable quantum theory of gravity
in four dimensions, Callan, Giddings, Harvey and Strominger
(CGHS) proposed a two dimensional renormalizable theory of
gravity in order to consider the problem of the endpoint of
a black hole \cite{CGHS}.
In their model, two-dimensional gravity is coupled with
a dilaton field $\phi$ and $N$ conformal matters
(free massless bosons) $f_i$:
\begin{equation}
\label{Action}
S_c = {1 \over 2\pi} \int \sqrt{-g} d^2x\, \Bigl[
{\rm {\rm e}}^{-2\phi} \Bigl\{ R + 4(\nabla \phi)^2 + 4\lambda^2 \Bigr\}
-{1 \over 2} \sum_{i=1}^N ( \nabla f_i )^2 \Bigr]
\end{equation}
They have shown that this model contains a black hole as a classical
solution and has a characteristic feature of the Hawking radiation by
including the trace anomaly from the conformal matters as
a quantum correction.
After that, many works have been performed in order to understand
this model \cite{RST}-\cite{Eli}.

In this paper, we show that this model also has a Lorentzian
wormhole as a classical solution. When we incorporate matters
in a form of a shock wave into the wormhole background, the structure
of the spacetime is changed into that of a black hole where
the curvature singularity is cloaked by a global event horizon.
Furthermore, we analyze the quantum correction
to the wormhole solution by using de Alwis' procedure \cite{deA},
and find that the classical wormhole solution would be stable against
quantum corrections.

We start with the review of the classical analysis of the CGHS
model \cite{CGHS}. After fixing the gauge symmetry in
the conformal gauge:
\begin{equation}
\label{ConformalGauge}
g_{\pm\mp} =  -{1 \over 2} {\rm e}^{2\rho}\   , \quad
g_{\pm\pm} =0 \ ,
\end{equation}
where $x^\pm = t \pm x$, the equations of motion and constraint equations
are given by
\begin{eqnarray}
\label{Tpmmp}
0&=&T_{\pm\mp} = - \partial_+ \partial_-({\rm e}^{-2\phi}) - \lambda ^2
{\rm e}^{-2\phi + 2\rho}\ ,\\
\label{dilaton}
0&=&-4 \partial_+\partial_-{\phi} + 4 \partial_+{\phi}\partial_-{\phi}
+2 \partial_+\partial_-{\rho} + {\lambda}^2 {\rm e}^{2\rho}\ ,\\
0&=&\partial_+\partial_-f_i \ ,\\
0&=&T_{\pm\pm} = {\rm e}^{-2\phi} ( 4 \partial_\pm {\rho} \partial_\pm {\phi}
- 2 \partial^2 _\pm {\phi} )
+ {1 \over 2}  \sum_{i=1}^N \partial_\pm f_i \partial_\pm f_i\ .
\end{eqnarray}
The general solution is obtained by CGHS as follows,
\begin{eqnarray}
\label{Soli}
{\rm e}^{-2\phi} &=& u - {\lambda}^2 \int^{x^+} {\rm e}^{w_+}
\int^{x^-} {\rm e}^{w_-} \ ,\\
\label{Solii}
{\rm e}^{-2\rho} &=& {\rm e}^{-w} {\rm e}^{-2\phi} \ ,\\
\label{Soliii}
f &=& f_+ (x^+) + f_- (x^-)\ .
\end{eqnarray}
Here $w$ is a sum of holomorphic and anti-holomorphic functions:
\begin{equation}
\label{Soliv}
w = w_+ (x^+) + w_- (x^-)\ ,
\end{equation}
and $u = u_+ (x^+) + u_- (x^-)$ is given by
\begin{equation}
\label{Solvi}
u_\pm = {M \over 2 \lambda} - {1 \over 2} \int {\rm e}^{w_\pm}
\int {\rm e}^{-w_\pm}
\sum_{i=1}^N \partial_\pm f_i \partial_\pm f_i.
\end{equation}
When $f_i=0$,
Callan {\it et al.} \cite{CGHS} have found that a solution
describing a black hole can be obtained by,
\begin{equation}
\label{BH}
{\rm e}^{-2\phi} = {\rm e}^{-2\rho}
                 = {M \over \lambda} - {\lambda}^2 x^+ x^-,
\end{equation}
if we fix the residual gauge symmetry by setting $w = 0$.

We can also obtain a Lorentzian wormhole solution by choosing the
following residual gauge fixing condition:
\begin{equation}
\label{WH}
{\rm e}^w = -4 x^+ x^-\ .
\end{equation}
Then by using Eqs.(\ref{Soli}) and (\ref{Solii}),
we find a solution is given by
\begin{eqnarray}
\label{WHi}
{\rm e}^{-2\phi} &=&
{M \over \lambda} + \lambda^2 (x^+ x^-)^2\ ,\nonumber \\
{\rm e}^{-2\rho} &=&  -{1 \over 4 x^+ x^-}{\rm e}^{-2\phi}\ .
\end{eqnarray}
As we will see later, this solution describes a Lorentzian wormhole.

Now we have several remarks. First, you might wonder
why the wormhole solution (\ref{WHi}) can be obtained by choosing the
gauge condition different from that of the black hole solution (\ref{BH}).
Usually, physics is not supposed to depend on the gauge choice.
This statement is, however, too naive.
As we will see soon, the geometrical and topological structures
of the spacetime in the black hole solution (\ref{BH}) and
the solution (\ref{WHi}) is distinct from each other.
Therefore we cannot transform the solution (\ref{WHi}) into
the black hole solution (\ref{BH}) by successive infinitesimal
coordinate transformations.

The second remark is that there does not appear any curvature singularity.
Usually curvature singularity appears when ${\rm e}^{-2\phi}$ vanishes
but ${\rm e}^{-2\phi}$ is always positive definite in
the solution (\ref{WHi}).

The apparent horizon which is defined by
$( \nabla {\rm e}^{-2\phi} )^2 = 0$ is located at $x^+ x^- =0$.
Since ${\rm e}^{-2\phi}$ and ${\rm e}^{-2\rho}$ should be positive
in the physically relevant spacetime regions, we find
\begin{equation}
\label{Penrose}
x^+ x^-<0 \ .
\end{equation}
Note that one cannot extend the present spacetime beyond
the apparent horizons $x^\pm=0$ without violating analyticity.
Therefore the Penrose diagram of this spacetime is composed of
only two asymptotically flat regions bounded by the event horizon
$x^\pm = 0$ and the spacetime infinities
$x^\pm \rightarrow \pm \infty$ (Fig.1).
Also note that the two asymptotically flat portions of this spacetime
are causally disconnected with each other.
If we regard the dilaton gravity theory as a theory which is obtained
from four dimensional gravity theory by the dimensional reduction,
we can identify the dilaton field with the radial coordinate,
$r^2={\rm e}^{-2\phi}$. Since ${\rm e}^{-2\phi}={M \over \lambda}$
when $x^\pm=0$, we can interpret the solution (\ref{WHi}) as
a solution describing a four dimensional spacetime which has a throat
of the radius $r^2 _H = {M \over \lambda} $.
The throat connects two asymptotically flat regions \cite{MTW}.
This is the reason why we call the structure of the spacetime
as ``the Lorentzian wormhole".
It should be noted that since there do not appear the regions which have
a black hole and/or a white hole singularity, the Lorentzian wormhole is
stationary and stable under the time development, which should be
contrasted with ``Schwarzshild wormhole" in four dimensions \cite{MTW}.

\unitlength 1mm

\begin{picture}(0,80)
\put(10,30){\line(1,1){20}}
\put(10,30){\line(1,-1){20}}
\put(30,50){\line(1,-1){40}}
\put(30,10){\line(1,1){40}}
\put(70,10){\line(1,1){20}}
\put(70,50){\line(1,-1){20}}
\put(30,30){$I'$}
\put(70,30){$I$}
\put(40,40){\line(1,2){8}}
\put(60,40){\line(-1,2){8}}
\put(45,60){horizons}
\put(80,40){\line(1,2){6}}
\put(86,52){\ future null infinity}
\put(80,20){\line(1,-2){6}}
\put(86,8){\ past null infinity}
\end{picture}

\noindent
{\bf Fig.1}
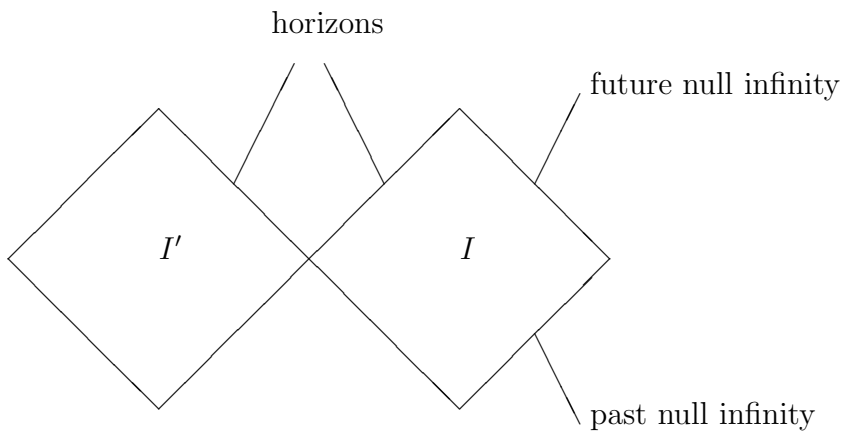
{\it The Penrose diagram of the Lorenzian wormhole.
The diagram is composed of two asymptotically
flat regions $I$ and $I'$ and there do not appear the regions which have
a black hole and/or a white hole singularity.}

\vskip 1cm

The third remark is that we do not have a {\it Euclidean} wormhole
as a classical solution in the CGHS model when we Wick-rotate the
signature of the metric.
Since there exist identities
$\partial_z {1 \over \bar z} = \partial_{\bar z} {1 \over z} = \pi \delta (z)$
in case of the Euclidean signature,\footnote{
The function ${1 \over z}$ is usually defined by
$\lim_{\epsilon\rightarrow 0}{1 \over z+\epsilon z\bar z}$ in the
neighbourhood of $z=0$. This definition, however, obviously violate
the analyticity.}
$w$ which is obtained by replacing $x^\pm$ with $z$, $\bar z$
in Eq.(\ref{WH}),
\begin{equation}
\label{WHE}
{\rm e}^w = -4 z\bar z\ ,
\end{equation}
does not satisfy the field equation
$\partial_z  \partial_{\bar z} w = 0$, which is obtained from the equations
corresponding to Eqs.(\ref{Tpmmp}) and (\ref{dilaton}).
($\partial_z  \partial_{\bar z} w
= \partial_z  \partial_{\bar z} {\rm ln}z
= \partial_z {1 \over \bar z}\neq 0$.)
Therefore $w$ in Eq.(\ref{WHE}) cannot be
given by a sum of analytic and anti-analytic functions as in Eq.(\ref{Soliv}).
This tells that we cannot choose the gauge fixing
conditions corresponding to Eq.(\ref{WH}).
On the other hand, in case of the Lorentzian signature,
we can consistently define ${ 1 \over x^\pm }$
in terms of the Cauchy principal value ${1 \over x^\pm}\rightarrow
\lim_{\epsilon\rightarrow 0}{1 \over x^\pm + i\epsilon}$.
without violating (anti-)holomorphic property.
We take the limit $ \epsilon \rightarrow 0$ at the final stage of
calculations in order to satisfy the Eq.(\ref{Soliv}).

Final remark is that we can choose a more general gauge
condition instead of Eq.(\ref{WH}),
\begin{equation}
\label{GeneralWH}
{\rm e}^w = -4 m^2 ( x^+ x^-)^{2m-1}  ,
\end{equation}
where $m$ is a positive integer. It is, however, easy to show that
the obtained solution can be transformed into the Eqs.(\ref{WHi})
by a nonsingular coordinate transformation.

Furthermore, there is another class of gauge conditions
which fix the residual gauge symmetry,
where ${\rm e}^w$ is given by the inverse powers of $x^+ x^-$,
\begin{equation}
\label{deSitterWH}
{\rm e}^w = {-4 \over (x^+ x^-)^3}\ .
\end{equation}
Then we obtain the following solutions,
\begin{eqnarray}
\label{deSitterWHi}
{\rm e}^{-2\phi}
&= {M \over {\lambda}} + {{\lambda}^2 \over (x^+ x^-)^2}\ ,\nonumber \\
{\rm e}^{-2\rho} &=  -{(x^+ x^-)^3 \over 4} {\rm e}^{-2\phi}\ .
\end{eqnarray}
The corresponding Penrose diagram has almost the same structure as
that of the previous wormhole in Fig.1 except that
two asymptotically de Sitter regions are connected with each other
in the solution (\ref{deSitterWHi}).
Note that the solution (\ref{deSitterWHi}) is transformed into
(\ref{WHi}) by the singular coordinate transformation
$x^\pm \rightarrow {1 \over x^\pm}$.
By choosing other complicated gauge conditions,
we can also construct a variety of solutions describing spacetimes
which have a complicated geometrical structure.

Now we consider the perturbation of the Lorentzian wormhole solution
Eq.(\ref{WHi}) by the matter field $f_i$. By following CGHS,
we consider the shock wave which travels in the $x^-$ direction
with magnitude $m$:
\begin{equation}
\label{Shock}
{1 \over 2}  \sum_{i=1}^N \partial_+ f_i \partial_+ f_i =
 m \delta ( x^+ -x^+ _0 ).
\end{equation}
Under the gauge condition (\ref{WH}), we find
\begin{eqnarray}
\label{ShockWHi}
{\rm e}^{-2\phi} &=& {M \over {\lambda}} + {\lambda}^2 (x^+ x^-)^2
-{m \over 2 x^+ _0} \Bigl\{ (x^+)^2 - (x^+ _0)^2 \Bigr\}
\theta(x^+ - x^+ _0)\ ,\nonumber \\
{\rm e}^{-2\rho} &=&  -{1 \over 4 x^+ x^-}{\rm e}^{-2\phi}\ .
\end{eqnarray}
When $x^+ < x^+ _0$, this solution describes the Lorentzian wormhole
connecting two asymptotically flat regions. When $x^+ > x^+ _0$,
this solution is identical with a black hole which has
a global event horizon
$x^- = x^- _0 \equiv -\sqrt{m \over {2\lambda}^2 x^+ _0}$
and the curvature singularity when ${\rm e}^{-2\phi} = 0$ (Fig.2).
This tells that the classical wormhole is unstable
against the matter perturbation.

\begin{picture}(0,80)
\put(0,30){\line(1,1){20}}
\put(0,30){\line(1,-1){20}}
\put(20,50){\line(1,-1){40}}
\put(20,10){\line(1,1){30}}
%%shock wave
\put(50,40){\line(1,-1){20}}
\put(49.5,39.5){\line(1,-1){20}}
%%curvature singularity
\put(50,40){\line(1,0){20}}
\put(50.2,39.8){\line(1,0){19.6}}
\put(50.5,39.5){\line(1,0){19}}
\put(70,40){\line(1,-1){10}}
\put(70,40){\line(-1,-1){10}}
\put(60,10){\line(1,1){20}}
\put(60,40){\line(1,2){5}}
\put(65,50){\ curvature singularity}
\put(65,35){\line(1,0){20}}
\put(85,35){\ shifted future horizon}
\put(65,25){\line(1,0){20}}
\put(85,25){\ matter shock wave}
\end{picture}

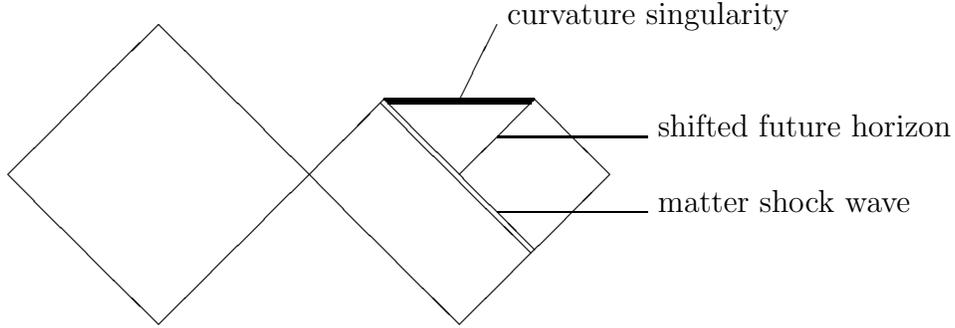
{\bf Fig.2} {\it The Penrose diagram where a matter shock wave collapses
into a wormhole. The shock wave creates a black hole.}

\vskip 1cm

We now consider the quantum theory of the wormhole.
By following de Alwis \cite{deA}, we define
the quantum fields $X$ and $Y$ as follows,
\begin{eqnarray}
\label{DefXY}
X &=& 2 \sqrt{2 \over |\kappa|} \int^\phi d \phi \  {\rm e}^{-2\phi}
\Bigl( 1 + {\kappa^2 \over 4} {\rm e}^{4\phi} \Bigr)^{1 \over 2}\ ,
\nonumber \\
Y &=&  \sqrt{2 |\kappa| } \Bigl( \rho - {1 \over \kappa}
{\rm e}^{-2 \phi} - \phi \Bigr)\ .
\end{eqnarray}
Here $\kappa={24-N \over 6}$ and $N$ is the number of
the free massless bosonic
matter fields.
The general solution of the equations of motion, which are given by the
quantum effective action has the following form:
\begin{equation}
\label{X}
X = - \sqrt{2 \over |\kappa|} u + {\lambda}^2 \sqrt{2 \over |\kappa|}
\int^{x^+} {\rm e}^{g_+}  \int^{x^-} {\rm e}^{g_-}
= -Y + \sqrt{|\kappa| \over 2} (g_+ + g_-) .
\end{equation}
Here $u$ is a sum of holomorphic and anti-holomorphic functions:
$u=u_+(x^+)+u_-(x^-)$ and the forms of $u_\pm$ are given in
Ref.\cite{deA}.
Note that $g_\pm$ corresponds to the degrees of the residual gauge
freedom which correspond to $w_\pm$ in the classical theory
in Eq.(\ref{Soliv}).
Therefore if we choose ${\rm e}^g = -4 x^+x^-$,
a quantum solution corresponding to the classical Lorentzian wormhole
solution (\ref{WHi})
\begin{eqnarray}
\label{QXY}
X &=& - \sqrt{2 \over |\kappa|}
\Bigl\{ u + \lambda^2 ( x^+x^-)^2 \Bigr\}\ ,\nonumber \\
Y &=& -X + \sqrt{ |\kappa| \over 2} \ln ( -4 x^+ x^- ) \ .
\end{eqnarray}
In the weak-coupling region $({\rm e}^{2 \phi} \ll 1 )$,
the above solution approaches to the classical solution,
\begin{eqnarray}
\label{WQWHi}
{\rm e}^{-2\phi} &\sim&  u + {\lambda}^2 (x^+ x^-)^2\ ,\nonumber \\
{\rm e}^{-2\rho} &\sim&  -{1 \over 4 x^+ x^-}{\rm e}^{-2\phi}\ .
\end{eqnarray}
On the other hand, in the strong-coupling region, we have
\begin{eqnarray}
\phi &\sim& {1 \over \kappa} \Bigl\{ u + {\lambda}^2 (x^+ x^-)^2
\Bigr\}\ ,\nonumber \\
{\rm e}^{2\rho} &\sim&  -4 x^+ x^- {\rm e}^{{2 \over \kappa}
{\rm e}^{-2 / \kappa} \Bigl\{ u + {\lambda}^2 (x^+ x^-)^2
\Bigr\}}\ .
\end{eqnarray}

In the solution (\ref{QXY}), there does not appear any curvature singularity.
By differentiating $X$ in the solution (\ref{QXY}) with respect to $x^\pm$,
we find that the apparent horizons are located at $x^+x^- = 0$.
Since the physically relevant region of spacetime is determined by
the the conditions that both ${\rm e}^{-2 \phi}$ and ${\rm e}^{-2 \rho}$
are positive, we find the regions $x^+x^- < 0$ are relevant even in
the above quantum wormhole solutions.
Therefore the structure of the spacetime in the quantum theory is
very similar to that in the classical theory.
This suggests that the classical
wormhole might be stable against the quantum effect.

In conclusion, we have found that there exists a solution which
describes the classical Lorentzian in two dimensional dilaton
gravity \cite{CGHS}.
Furthermore, we have examined the quantum corrections to the wormhole
solution by following the quantization procedure by de Alwis
and it has been shown that the wormhole solution is
stable against quantum correction and matter perturbation.
The spacetime in the wormhole solution consists of two asymptotically
flat regions and does not involve any region where a black hole or
a white hole singularity appears.
Even though the two asymptotically flat regions are causally disconnected
with each other, we would like to expect that informations
in one of the asymptotically flat region leaks into another one since
the wormhole is stable under time development \cite{MTW}.

Finally,
we have a comment on the relation with other works. The authors of
Ref.\cite{Hotta} also have analyzed the wormhole in the modified
CGHS model. We also have a solution analogous to their
wormhole solution in the present framework by
choosing the following residual gauge conditions:
\begin{eqnarray}
\label{Partition}
w_+ &= \lambda x^+ , \ \
w_- = - \lambda x^- - {1 \over 2} \ln ( 1- {\rm e}^{2 \lambda x^-} ),
\ \ \ &{\rm when} \ \ x^1 > 0 \nonumber \\
w_+ &= - \lambda x^+ - {1 \over 2} \ln ( 1- {\rm e}^{2 \lambda x^+} ), \ \
w_- = \lambda x^- \ \ \ &{\rm when} \ \ x^1 < 0\ .
\end{eqnarray}

\vskip 1cm

\noindent
{\bf Acknowledgement}

We are grateful to Professor T. Banks for valuable discussions.
This work was partly done when S.N. was in KEK, Theory Group
and I.O. in SLAC. We would like to thank these institutes for
kind hospitality. I.O. also wishes to acknowledge the financial
support of the Japan Society for the Promotion of Science and
SLAC.

\end{document}